%% file: main.tex
\documentclass[conference]{IEEEtran}
\IEEEoverridecommandlockouts
\usepackage{amsmath,amssymb,amsfonts}
\usepackage{newtxtext,newtxmath}
\usepackage{graphicx}
\usepackage{textcomp}
\usepackage{siunitx}
\usepackage{booktabs}
\usepackage{xcolor}
\usepackage{authblk}
\usepackage{algorithm}
\usepackage{algpseudocode}

\usepackage[sorting=none]{biblatex}
\def\BibTeX{{\rm B\kern-.05em{\sc i\kern-.025em b}\kern-.08em
    T\kern-.1667em\lower.7ex\hbox{E}\kern-.125emX}}
\bibliography{lit.bib}

\begin{document}
\title{PhaseGen: A Diffusion-Based Approach for Complex-Valued MRI Data Generation}
\author[1, 2, 4]{Moritz Rempe}
\author[1, 2, 4]{Fabian Hörst}
\author[1]{Helmut Becker} 
\author[4]{Marco Schlimbach}
\author[1, 7]{\\Lukas Rotkopf}
\author[4]{Kevin Kröninger}
\author[1, 2, 3, 4, 5, 6]{Jens Kleesiek}
\affil[1]{Institute for AI in Medicine (IKIM), University Hospital Essen, Girardetstraße 2, 45131 Essen, Germany}
\affil[2]{Cancer Research Center Cologne Essen (CCCE), University Medicine Essen, Hufelandstraße 55, 45147 Essen, Germany}
\affil[3]{RACOON Study Group, Site Essen, Essen Germany}
\affil[4]{Department of Physics, Technical University Dortmund, Otto-Hahn-Straße 4a, 44227 Dortmund, Germany}
\affil[5]{German Cancer Consortium (DKTK), Partner Site Essen, Hufelandstraße 55, 45147 Essen, Germany}
\affil[6]{Medical Faculty and Faculty of Computer Science, University of Duisburg-Essen, 45141 Essen, Germany}
\affil[7]{German Cancer Research Center (DKFZ), 69120 Heidelberg, Germany}
\setcounter{Maxaffil}{0}
\renewcommand\Affilfont{\itshape\small}

\maketitle 

\begin{abstract}
Magnetic resonance imaging (MRI) raw data, or k-Space data, is complex-valued, containing both magnitude and phase information. However, clinical and existing Artificial Intelligence (AI)-based methods focus only on magnitude images, discarding the phase data despite its potential for downstream tasks, such as tumor segmentation and classification. 

In this work, we introduce \textit{PhaseGen}, a novel complex-valued diffusion model for generating synthetic MRI raw data conditioned on magnitude images, commonly used in clinical practice. This enables the creation of artificial complex-valued raw data, allowing pretraining for models that require k-Space information. 

We evaluate PhaseGen on two tasks: skull-stripping directly in k-Space and MRI reconstruction using the publicly available FastMRI dataset. Our results show that training with synthetic phase data significantly improves generalization for skull-stripping on real-world data, with an increased segmentation accuracy from 41.1\% to 80.1\%, and enhances MRI reconstruction when combined with limited real-world data. 

This work presents a step forward in utilizing generative AI to bridge the gap between magnitude-based datasets and the complex-valued nature of MRI raw data. This approach allows researchers to leverage the vast amount of avaliable image domain data in combination with the information-rich k-Space data for more accurate and efficient diagnostic tasks.

We make our code publicly available at \url{https://github.com/TIO-IKIM/PhaseGen}.

\end{abstract}


\section{Introduction}

Magnetic resonance imaging (MRI) is one of the most common clinical imaging procedures. 
Due to its high resolution and ability to visualize soft tissue, MRI is used in a wide range of medical applications, including the diagnosis of cancer, neurological disorders, and musculoskeletal diseases.
With the advent of artificial intelligence (AI), new methods for improving or automating these medical applications are being developed \cite{eggerDeepLearningFirst2021a, egger2022medical, wang2025deep}.
While most of these methods are designed for the usage of magnitude image domain data \cite{albuquerque2025deep}, the initial MRI raw data is acquired in the so-called "k-Space", the frequency domain. This raw data is complex-valued, including magnitude and phase data. While this frequency domain data is not interpretable by humans, this additional data - when compared to the commonly used magnitude data in the image domain - provides neural networks with additional information. 
Until now, this raw data has primarily been used for reconstruction and transformation into the image domain and discarded afterwards. While recent work has demonstrated the benefits of using these complex-valued raw data in diagnostic tasks \cite{rempeTumorLikelihoodEstimation2024, liClassificationRegressionSegmentation2025}, research and publicly available datasets are scarce. Dishner et. al state a total of 110 MRI datasets containing 1,68 million individual subjects, with potential use for AI \cite{dishnerSurveyPubliclyAvailable2024}. This vast amount of data only includes the magnitude image domain data, with the additional phase information of complex-valued raw data being discarded. When looking at the available MRI k-Space datasets, there is only a handful of contributions, such as the popular "FastMRI" datasets \cite{zbontarFastMRIOpenDataset2018, tibrewala2024fastmri, solomon2025fastmri}, with roughly \SI{9300} subjects in total. Moreover, it should be noted that the publicly available raw data is exclusively intended for the purpose of reconstructing undersampled MRI data and so far not for diagnostic purposes. In contrast to image domain datasets, it is not intended for tasks such as classification or segmentation \cite{singhEmergingTrendsFast2023}. Due to the lack of available data, or necessary label for classification or segmentation tasks, there is a lot of untapped potential of MRI research. 
In the past few years, especially with the rise of diffusion models, the field of generative AI experienced a lot of attention. With the help of generative models, synthetic datasets can be created which improve model performance, by artificially expanding the available training data \cite{croitoruDiffusionModelsVision2023}. 

\begin{figure*}[htbp]
    \centerline{\includegraphics[width=\textwidth]{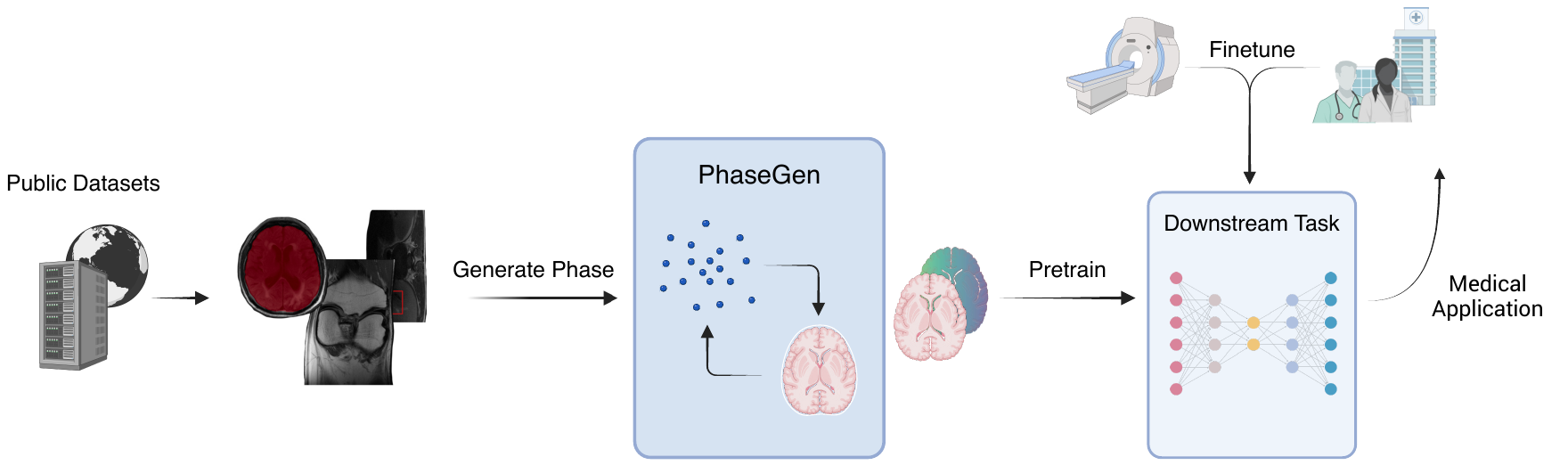}}
    \caption{Overview of the proposed method. Publicly available magnitude image domain data is used to generate synthetic complex-valued k-Space data. This synthetic data facilitates pretraining of models for clinical downstream tasks, which can later be fine-tuned using real-world data.}
    \label{fig:overview}
\end{figure*}

In this work we utilize a novel complex-valued diffusion model which can create synthetic MRI k-Space data, guided by the magnitude image domain data. With our model, called "PhaseGen", researchers and clinicians are now able to translate the vast amount of available image domain data into synthetic complex-valued raw data to use it for pretraining their models. Later fine-tuning with real-world data can then be conducted, to improve the model performance and making them applicable for clinical use. This approach comes with the benefit of reducing the need for real-world data, which is often scarce and hard to obtain. Focus can then be on less, but high quality data, which can be used for fine-tuning. As an example, Fig. \ref{fig:overview} shows the workflow for a possible application of our work.

We train our model on real-world clinical MRI raw data and evaluate it on two different tasks, skull-stripping in k-Space and MRI reconstruction on the publicly available FastMRI knee dataset \cite{knollFastMRIPubliclyAvailable2020}.
The code for our proposed method is publicly available at \url{https://github.com/TIO-IKIM/PhaseGen}.

\section{Related Work}

MRI raw data is mainly used for image processing steps, including coil combination and reconstruction. Datasets such as FastMRI Prostate \cite{tibrewalaFastMRIProstatePublicly2023} or Knee \cite{knollFastMRIPubliclyAvailable2020} have been collected to advance research in the field of MRI reconstruction. By undersampling the k-Space and then applying algorithms to fill the missing k-Space lines, the acquisition time of MRI scans can be reduced significantly. Many of these reconstruction algorithms are based on deep neural networks \cite{heckelDeepLearningAccelerated2024}. 
Besides reconstruction, the information-rich k-Space is mainly ignored for other downstream tasks. Li et al. \cite{liClassificationRegressionSegmentation2025} were able to show the benefits of utilizing the k-Space for diagnostic tasks in cardiac MRI. ``k-Strip" showed, that segmentations derived directly from the k-Space, without transformation into the image domain, is possible \cite{rempeKstripNovelSegmentation2024a}. Additionally, there are examples of pathologies which are detectable more easily with the help of the phase information in susceptibility-weighted imaging, such as thrombosis \cite{bradburyMesentericVenousThrombosis2002}, bleedings \cite{weng2020black} or calcification \cite{wuIdentificationCalcificationMRI2009}.

When working with the complex-valued k-Space, the question arises whether to split the complex values into real and imaginary channel or to use fully complex-valued neural networks. Complex valued neural networks (CVNNs) can represent the nature of this input data, making them more versatile in this domain \cite{leeComplexValuedNeuralNetworks2022}. Common use cases of CVNNs are signal processing, such as complex-valued sonar or radio frequencies \cite{smithComplexValuedNeuralNetworks2023}, as well as acoustic recognition algorithms \cite{hayakawaApplyingComplexValuedNeural2018}.

When training deep learning algorithms, the amount of high-quality data is essential. In recent years, new ways of artificially increasing the amount of available training data have been explored. The classical data augmentation approach is using image manipulations, such as geometric transformations, color mixing or random erasing \cite{shortenSurveyImageData2019}. While these approaches can have a great impact on the model performance, the underlying image information remain the same. Deep learning-based data augmentation approaches attempt to artificially increase the amount of available data by generating synthetic data which resembles real data.  
While variational autoencoders \cite{kingmaAutoEncodingVariationalBayes2013} were the most common generative models for a long time, the field of generative AI became more popular with the development of the concept of generative adversarial networks (GANs) \cite{goodfellowGenerativeAdversarialNetworks2020}. GANs utilize two seperate neural networks, one being the generative model and the other being the ``discriminator", rating the generated output in comparison to real data. While still frequently used for different generative tasks, such as the generation of artifical brain MRI images \cite{ferreiraImprovedMultiTaskBrain2024}, a novel approach, called ``diffusion probabilistic models" emerged. Trabucco et al. use a diffusion model to generate synthetic photography images for few-shot image classification and report an improvement of accuracy in multiple domains \cite{trabuccoEffectiveDataAugmentation2023}. 
Diffusion probabilistic models have also been applied to the task of MRI reconstruction, such as the unrolled diffusion model by Korkmaz et al. \cite{korkmazSelfsupervisedMRIReconstruction2023}.

While diffusion models are frequently used for data generation in different domains, to the best of our knowledge, there is currently no work on the generation of artificial complex-valued k-Space data. We present a complex-valued diffusion model, which presents a novelty in the field of MRI data generation. Additionaly, it contributes to the research of complex-valued neural networks, by presenting a diffusion model which works with complex-valued input data and noise. 

\section{Material and Methods}
\subsection{MRI raw data}

MRI raw data is acquired in k-Space, the frequency domain. These data are complex-valued, consisting of magnitude and phase. While the magnitude can take on arbitrary positive values, the phase is bound to the range of $-\pi$ to $\pi$. Via the inverse Fourier transformation, the k-Space data can be transformed into the complex-valued image domain. This transformation is fully reversible. Most downstream tasks only utilize the magnitude information, discarding the phase data. 

\subsection{Complex Valued Neural Networks}

Complex valued neural networks (CVNNs) are a type of neural network that can process complex numbers as input and output. They are designed to work with data that has both real and imaginary components, such as signals in the frequency domain. In CVNNs, the weights and activations are represented as complex numbers, allowing them to capture the phase information inherent in the data. This is particularly useful in applications like signal processing, where phase information is crucial for accurate analysis and reconstruction. There are two common approaches to implement CVNNs: the first one is to split the complex numbers into real and imaginary parts, treating them as separate channels. The second approach is to use complex-valued operations directly in the network architecture, allowing for more efficient processing of complex data. For further reading, we recommend the work by Hirose et al. \cite{hirose2006complex}. In this work we will use the second approach, as it allows us to fully utilize the complex nature of the input data.

\subsection{(Complex Valued) Diffusion \& Model Architecture}

\begin{figure}[!bp]
    \centerline{\includegraphics[width=\linewidth]{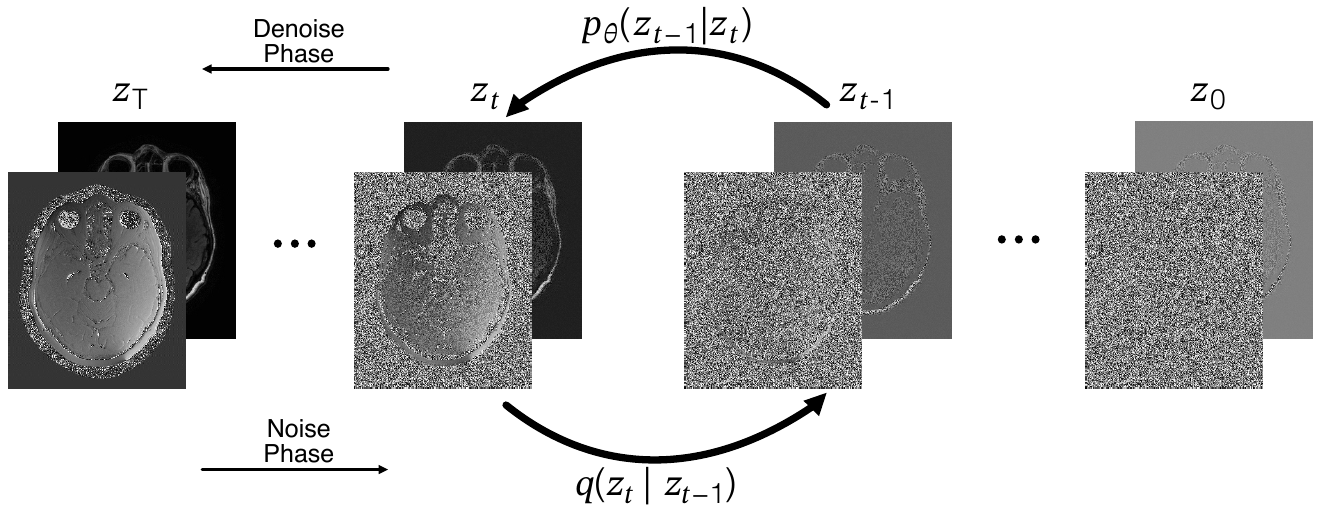}}
    \caption{Graphical representation of the complex-valued forward and reverse diffusion process. The input $z_t$ consists of magnitude and phase in the image domain. The added complex-valued noise primarily affects the phase while preserving magnitude.}
    \label{fig:fig_diffusion}
\end{figure}

The goal of diffusion models differs from other neural networks in that the network learns the diffusion process, introduced by gradually increasing noise to an input image. Given a data point $x_0$, in the \textit{forward diffusion process} during training, small amounts of Gaussian noise\footnote{There are works showing the benefits of other types of noise \cite{nachmaniNonGaussianDenoising2021}, but Gaussian noise is commonly used, as the training tends to be more stable.} is added to the image in $T$ consecutive time steps. The resulting data, superimposed with noise, $x_t$ can be described with
\begin{equation}
    q(x_t | x_{t-1}) = \mathcal{N}(x_t; \sqrt{1-\beta_t}\cdot x_{t-1}, \beta_t I), 
\end{equation} 
where $q$ is the forward diffusion process, $\mathcal{N}$ a normal distribution, $I$ the identitiy matrix and $\beta_t$ the scheduler, which defines the amount of noise $\epsilon$ added, commonly ranging from $0$ to $1$. During training, the model will then learn to predict the added noise in each timestep during the \textit{reverse diffusion process} of the diffusion model parametrized by $\theta$
\begin{equation}
    p_\theta(x_{t-1}|x_t) = \mathcal{N}(x_{t-1}; \mu_\theta(x_t, t), \Sigma_\theta(x_t, t)),
\end{equation} 
with $\mu$ being the mean $\sqrt{1-\beta_t} \cdot x_{t-1}$.
To update the weights of the model, the loss is calculated for each timestep. 
During inference, the diffusion model then takes the normal noise distribution as input and via reverse diffusion samples a new data point. 

Because MRI raw data is complex-valued, we have to adapt the diffusion and sampling process. With the magnitude already being present in the data, we only want to generate a corresponding phase. To simulate the nature of complex values $z$ we define a complex-valued noise distribution $\mathcal{N}_z$. To focus on the phase to be generated, we adapt $\theta$ to have a magnitude of one, while the phase values are in the normal distribution $\mathcal{U}$ with mean $0$ and variance $1$:
\begin{equation*}
    \mathcal{N}_z = e^{i\epsilon}; \ \ \epsilon \backsim \mathcal{U}(-\pi, \pi)  .  
\end{equation*}  
The complex-valued forward diffusion step can then be calculated as
\begin{equation}
    z_{t} = (|z_{t-1}| \sqrt{\alpha_t} + |\epsilon|\sqrt{1-\alpha_t}) \cdot \text{exp}(\angle z_{t-1} + \angle \epsilon\sqrt {1-\alpha_t})
\end{equation}
with $\alpha_t:=1-\beta_t$, |.| denoting the magnitude and $\angle$ the phase of a complex number.
A graphical representation of the complex-valued forward and reverse diffusion process can be seen in Fig. \ref{fig:fig_diffusion}. The corresponding algorithms for training and sampling are depicted in Alg. \ref{fig:algorithms}.
As underlying model architecture we use a complex-valued residual U-Net structure, adapted from the "k-Strip" algorithm \cite{rempeKstripNovelSegmentation2024a}. The model takes as input during training $z_t$, to which noise is gradually added and in a second channel the magnitude of the input, which is supposed to be consistent throughout the process. During inference, the input consists of randomly sampled noise $\mathcal{N}_z$ and the magnitude in a second channel. The second channel is kept constant in every sampling step.

\renewcommand{\figurename}{Alg.}
\setcounter{figure}{0} 
\begin{figure}[!tbp]
    \centering
    \begin{algorithm}[H]
        \caption{Training}\label{alg:train}
        \begin{algorithmic}[1]
            \State \textbf{repeat}
            \State \ \ $z_0 \backsim q(z_0)$
            \State \ \ $t \backsim \text{Uniform}(\{1, ..., T\})$
            \State \ \ $\epsilon \backsim \mathcal{N}_z$
            \State \ \ Take gradient descent step on \\
             \ \ \ \ \ \ $\bigtriangledown_\theta ||\epsilon - \epsilon_\theta(\sqrt{\bar{\alpha}}z_0 + \sqrt{1-\bar{\alpha}_t}\epsilon, t)||^2$
            \State \textbf{until} convergence
        \end{algorithmic}
    \end{algorithm}
    \begin{algorithm}[H]
        \caption{Sampling}\label{alg:sample}
        \begin{algorithmic}[1]
            \State $z_T \backsim \mathcal{N}_z$
            \State \textbf{for} $t = T, ..., 1$ \textbf{do}
            \State \ \ $\eta \backsim \mathcal{N}(0, \text{I})$, if t > 0, else $\eta = 0$
            \State \ \ $z_{t-1} = \frac{1}{\sqrt{\alpha_t}}\left(z_t - \frac{1-\alpha_t}{\sqrt{1-\bar{\alpha_t}}}\epsilon_\theta(x_t, t) \right) + \sigma_t\eta$
            \State \textbf{end for}
            \State \textbf{return} $z_0$
        \end{algorithmic}
    \end{algorithm}
    \caption{Algorithms for training and sampling in the complex-valued diffusion model. Adapted from \cite{hoDenoisingDiffusionProbabilistic2020a}.}
    \label{fig:algorithms}
\end{figure}
\renewcommand{\figurename}{Fig.}
\setcounter{figure}{2} 

To validate the proper phase generation of our proposed method, we will conduct multiple experiments. The first downstream task is skullstripping directly in the k-Space. This segmentation task, which separates the brain tissue from the skull in an image (in this case directly in the frequency domain), has already been presented in \cite{rempeKstripNovelSegmentation2024a}. The accuracy is tested on a raw dataset gathered at the University Hospital Essen. The comparison will be conducted on the model trained with artifical phase information created with the presented approach and other methods for data generation, such as random sampling.

In the second experiment, we will use the artifical raw data to train a reconstruction algorithms on the publicly available FastMRI knee single coil dataset.

\subsection{Datasets}

For the training of our diffusion model, we use a raw dataset gathered at the University Hospital Essen [24-11872-BO] on two seperate MRI machines (Siemens 1.5T and 3T). In total we use \SI{12071} 2D raw data scans from 390 patients, consisting of different $T_1$ and $T_2$ sequences, as well as different resolutions. Each slice comprises $256\times256$ pixels.

For the first validation experiment we use a dataset from the University Hospital Essen, containing \SI{21822} 2D brain images from 150 patients, scanned with a 1.5 T and 3 T MRI machine, to train the skullstripping model. Some of these scans contain pathologies. Corresponding brain masks are already available as ground truth. Because this dataset only consists of magnitude image domain scans, we generate the phase data with our presented method and transform the data into the k-Space via the Fourier transformation. To investigate the benefit of training with artificially generated MRI raw data, we validate the trained model on real-world raw data, in total 14 volumes of individual patients, scanned on the two MRI machines mentioned above. The ground truth in the image domain is generated with the STAPLE algorithm \cite{warfieldSimultaneousTruthPerformance2004a}, combining the results of three different skullstripping algorithms HD-BET \cite{isenseeAutomatedBrainExtraction2019b}, Synthstrip \cite{hoopesSynthStripSkullstrippingAny2022a} and the de-identification tool presented in \cite{rempeDeIdentificationMedicalImaging2024}, on the image domain magnitude data.

In the second validation experiment, performing the MRI reconstruction, we use the FastMRI single coil knee dataset for training and testing. 

An overview of the used datasets can be found in Appendix \ref{tab:datasets}.

\subsection{Diffusion model training}

For training, the Adam optimizer is used with an initial learning rate of 1e-4. The learning rate is reduced exponentially with a gamma factor of 0.995, a beta coefficient of 0.99 and
an epsilon value of 1e-08. After each encoder convolution a dropout of 20\% is chosen. The model is trained for 200 epochs with a batch size of 128. The overall training time takes 10 hours on an NVIDIA A100 GPU with 80GB of graphics memory. The diffusion model uses a cosine noise scheduler with 1000 timesteps and an $\epsilon$ of 0.008. The model has a total of 30.4 million parameters.

\section{Experiments \& Results}
\subsection{PhaseGen}
\begin{figure*}[!htb]
    \centering
    \includegraphics[width=15cm]{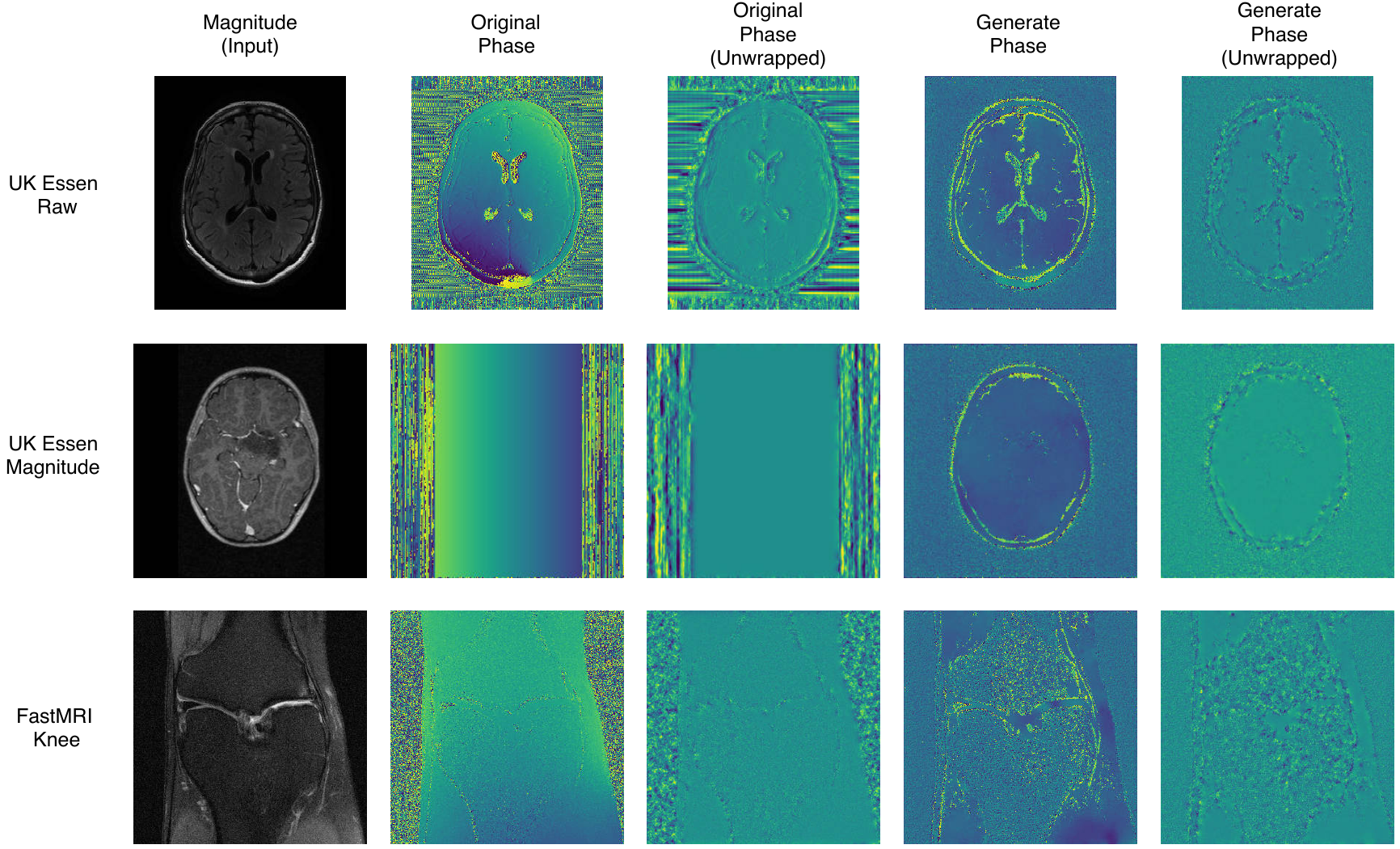}
    \caption{Example outputs of PhaseGen compared to original data. From left to right: input magnitude image, corresponding original phase, PhaseGen-predicted phase, unwrapped original phase, unwrapped predicted phase.}
    \label{fig:exampleDiff}
\end{figure*}

Exemplary results of the phase generation with the proposed method can be seen in Fig. \ref{fig:exampleDiff}. The model does not alter the magnitude input data, but generates a corresponding phase. Because the original phase is also dependent on information not available to the model, such as the coil sensitivity, the generated phase is not identical to the original phase. Nevertheless, the generated phase is consistent with the magnitude data and resembles possible phase images. To further understand the quality of the generated phase, additional phase unwrapping was performed on the generated, as well as the original phase. The used algorithm is based on a publicly available \textit{Laplacian-Based Phase Unwrapping tool}\footnote{\url{https://github.com/blakedewey/phase_unwrap}} \cite{dewey2022laplacian}.

\subsection{Skullstripping}

Skullstripping is the task of removing the skull from a brain image, only leaving the brain tissue. This segmentation task is commonly used in the image domain on the magnitude data. In \cite{rempeKstripNovelSegmentation2024a} we showed the feasability of performing skullstripping directly in the k-Space, preserving valuable phase information for further downstream tasks. While the model performed well on synthetic datasets, generalizing to real-world clinical data proved challenging. In this first experiment we show the benefits of generating artifical phase for magnitude training data, leading to superior generalization on real-world raw data. The same model is trained on the magnitude image domain data and the artifical phase data, generated with the proposed complex-valued diffusion model. The model is then tested on real-world clinical data with original-phase values, to validate the benefit of training with the artifical phase data. 

The following data generation methods are compared:
\begin{itemize}
    \item No phase data
    \item Naive phase generation
    \item Phase generation with the proposed diffusion model
\end{itemize}

The naive phase data generation, creates a synthetic phase by superimposing sinusoidal functions along both spatial dimensions and modulating it with the normalized magnitude to maintain anatomical structure correlation. Small random variations ($\sigma = 0.05$) were added to introduce realistic phase noise. The resulting synthetic phase can be described as 
\begin{equation}
    \phi(x,y) = [\sin\left(\frac{2\pi x}{N}\right) + \cos\left(\frac{2\pi y}{N}\right)] \cdot \hat{M}(x,y) + \eta(x,y) ,
\end{equation}
where $N$ is the image size, $\hat{M}$ the normalized magnitude and $\eta$ the added noise.

\begin{figure}[!hb]
    \centering
    \includegraphics[width=8.5cm]{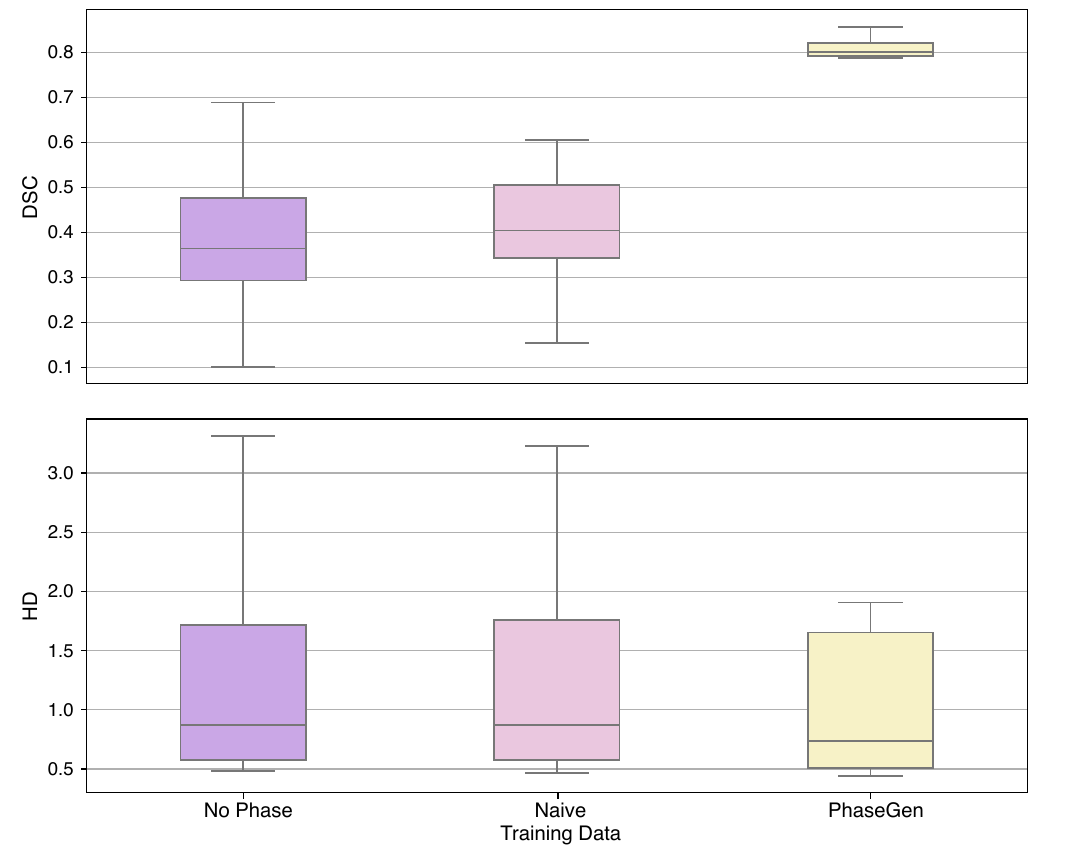}
    \caption{Comparison of different phase data generation methods on the task of skullstripping directly in the frequency domain. The model trained with the data generated by the proposed methods outperforms the other methods in both metrics, DSC and HD, showing superior segmentation performance.}
    \label{fig:boxplot_skullstrip}
\end{figure}

\begin{table}[!htb]
    \caption{Results of the skullstripping task. Compared are the three training data generation methods.}
    \centering
    \def\arraystretch{1.6}
    \begin{tabular}{lll}
    \toprule
    {Training Data} & {DSC ($\%) \uparrow$} & {HD (pixel) $\downarrow$}\\
    \midrule
    No Phase & 40.1 ± 17.2 & 1.577 ± 1.799\\
    Naive & 41.1 ± 13.5 & 1.634 ± 1.946\\
    Diffusion (Proposed) & \textbf{80.1 ± 3.2} & \textbf{1.534 ± 1.923} \\
    \bottomrule
    \end{tabular}
    \label{tab:1}
\end{table}

For each method, the resulting data is transformed into the k-Space. For all three methods, an extensive grid search was performed to find the best hyperparameters. An overview of the used hyperparameters can be found in Appendix \ref{tab:hyperparameters_skullstrip}. The model architecture, a complex-valued residual U-Net, as presented in \cite{rempeKstripNovelSegmentation2024a}, is the same for all three methods. Early stopping is applied as soon as the validation loss does not improve for 30 consecutive epochs.
The results are shown in Fig. \ref{fig:boxplot_skullstrip} and Tab. \ref{tab:1}. The proposed method outperforms the other methods in both metrics, Dice similarity coefficient (DSC) and Hausdorff distance (HD). The model trained with the proposed method achieves a DSC of 80.1\% and a HD of 1.534 pixel. The model trained with naive phase data achieves a DSC of 41.1\% and a HD of 1.634 pixel. The model trained without phase data achieves a DSC of 40.1\% and a HD of 1.577 pixel.
It has to be noted, that the large difference in DSC is based on the fact, that both the model trained with no phase and the model with naive phase generation, are not able to generalize on the real-world data.

\subsection{MRI Reconstruction}

\begin{figure*}[!tb]
    \centering
    \includegraphics[width=\textwidth]{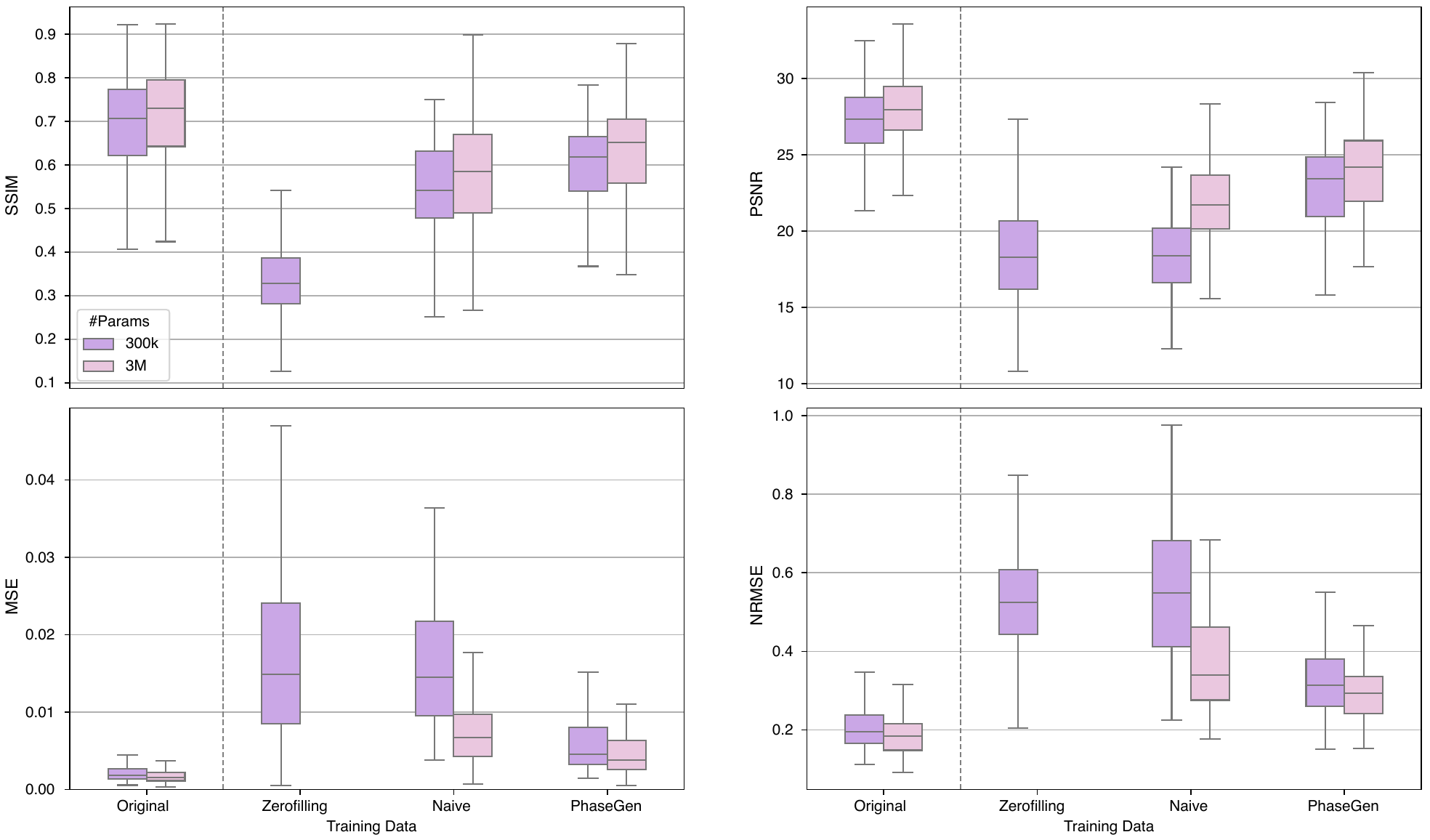}
    \caption{Results of the reconstruction task. Shown are the SSIM, PSNR, MSE and NRMSE for the model trained with original-phase data, naive phase generation and phase generation with the proposed method, as well as zerofilling as the baseline method. The results are shown for an undersampling factor of four. Higher values in the top row are better, while lower values in the bottom row are better.}
    \label{fig:results_recon_1}
\end{figure*}

The second validation experiment is the MRI reconstruction task. 
To reduce MRI acquistion times, the k-Space is often undersampled. This means that only a small part of the k-Space is acquired, while the rest is filled with zeros. This leads to a loss of information and thus to a lower image quality. The goal of this task is to reconstruct the missing k-Space lines, using the available data \cite{hyun2018deep}.

The FastMRI singlecoil knee dataset is used for training and testing a complex-valued residual U-Net, based on the architecture described above. We extend the existing model with data-consistency layers in the downsample path. We compare the model results of the model trained with the original raw data with the model trained on artificially generated phase data with the proposed method. The results are again compared with a model trained on naive phase data, generated as explained above. The used hyperparamters can be found in Appendix \ref{tab:hyperparameters_reconstruction}. All experiments are conducted with a small model with roughly \SI{209000} parameters (209k model) and a larger model with 3.3 million parameters (3M model). Following the official FastMRI challenge implementations\footnote{\url{https://github.com/facebookresearch/fastMRI}}, we use Cartesian undersampling masks with an 8\% fully sampled center region for an undersampling rate of four and a 4\% region for an undersampling rate of eight.

To evaluate the performance of the reconstruction we use the commonly used metrics Peak Signal to Noise Ratio (PSNR) and Structural Similarity Index (SSIM), as well as the Mean Squared Error (MSE) and the Normalized Root Mean Squared Error (NRMSE). The results are shown in Tab.\ref{tab:3} and Fig.\ref{fig:results_recon_1}. 

\begin{table}[!tb]
    \caption{Results of the reconstruction task. Compared is the same model with different sources of training data: naive phase generation, phase generation with PhaseGen and original-phase data, with zerofilling as the baseline. The numbers inside the bracket indicate the number of parameters of the model. The results are shown for an undersampling factor of four and eight.}
    \centering
    \def\arraystretch{1.2}
    \resizebox{\columnwidth}{!}{%
    \begin{tabular}{lllll}
    \toprule
    \multicolumn{1}{c}{Training Data} & \multicolumn{1}{c}{SSIM ($\%) \uparrow$} & \multicolumn{1}{c}{PSNR (dB) $\uparrow$} & \multicolumn{1}{c}{MSE $\downarrow$} &  \multicolumn{1}{c}{NRMSE $\downarrow$}\\
    \midrule
    Undersampling x4 & & & & \\
    \hspace{0.8cm} \textit{Original [209k]} & \textit{69.10 ± 11.44} & \textit{27.19 ± 2.22} & \textit{0.002 ± 0.001} & \textit{0.207 ± 0.061}  \\
    \hspace{0.8cm} \textit{Original [3.3M]} & \textit{71.28 ± 11.76} & \textit{27.98 ± 2.58} & \textit{0.002 ± 0.001} & \textit{0.190 ± 0.066}  \\
    \hspace{0.8cm} Zerofilling & 33.88 ± 7.89 & 18.62 ± 3.32 & 0.017 ± 0.011 & 0.553 ± 0.153 \\
    \hspace{0.8cm} Naive [209k] & 54.32 ± 10.64 & 18.31 ± 2.53 & 0.018 ± 0.011 & 0.606 ± 0.289 \\
    \hspace{0.8cm} Naive [3.3M] & 56.31 ± 14.44 & 21.81 ± 3.01 & 0.008 ± 0.007 & 0.403 ± 0.192 \\
    \hspace{0.8cm} \textbf{PhaseGen [209k]} & \textbf{59.73 ± 9.93} & \textbf{22.76 ± 2.94} & \textbf{0.007 ± 0.007} & \textbf{0.351 ± 0.134} \\
    \hspace{0.8cm} \textbf{PhaseGen [3.3M]} & \textbf{63.16 ± 10.87} & \textbf{23.95 ± 2.91} & \textbf{0.005 ± 0.004} & \textbf{0.301 ± 0.098} \\
    \midrule
    Undersampling x8 & & & & \\    
    \hspace{0.8cm} \textit{Original [209k]} & \textit{64.74 ± 11.38} & \textit{25.58 ± 2.46} & \textit{0.003 ± 0.002} & \textit{0.247 ± 0.069} \\
    \hspace{0.8cm} \textit{Original [3.3M]} & \textit{67.01 ± 11.97} & \textit{26.26 ± 2.61} & \textit{0.003 ± 0.002} & \textit{0.231 ± 0.071} \\
    \hspace{0.8cm} Zerofilling & 31.15 ± 8.47 & 17.12 ± 3.58 & 0.025 ± 0.016 & 0.673 ± 0.248 \\
    \hspace{0.8cm} Naive [209k] & 55.74 ± 11.11 & 20.41 ± 3.26 & 0.012 ± 0.008 & 0.465 ± 0.199 \\
    \hspace{0.8cm} Naive [3.3M] & 47.52 ± 13.41 & 20.32 ± 2.57 & 0.011 ± 0.007 & 0.454 ± 0.139 \\
    \hspace{0.8cm} \textbf{PhaseGen [209k]} & \textbf{54.57 ± 11.53} & \textbf{20.61 ± 3.28} & \textbf{0.011 ± 0.009} & \textbf{0.434 ± 0.096} \\
    \hspace{0.8cm} \textbf{PhaseGen [3.3M]} & \textbf{55.39 ± 12.11} & \textbf{21.22 ± 3.22} & \textbf{0.010 ± 0.010} & \textbf{0.405 ± 0.096} \\
    \bottomrule
    \end{tabular}}
    \label{tab:3}
\end{table}

The model with 3.3 million parameters trained with original-phase data achieves a PSNR of 27.98 dB and an SSIM of 71.28\% for an undersampling factor of four, while the model trained with PhaseGen achieves a PSNR of 23.95 dB and an SSIM of 63.16\%. The model trained with naive phase generation achieves a PSNR of 21.81 dB and an SSIM of 56.31\%. While the model trained with the artifical data generated by PhaseGen is able to reach better scores than the model trained with naive phase generation, it is not able to reach the same results as the model trained with original-phase data. These results show, that the proposed model can generate data which better resembles original-phase data than naive phase generation, but is not able to reach the same results as the model trained with original-phase data, especially when used at higher undersampling rates. 

\begin{figure}[!tb]
    \centering
    \includegraphics[width=\linewidth]{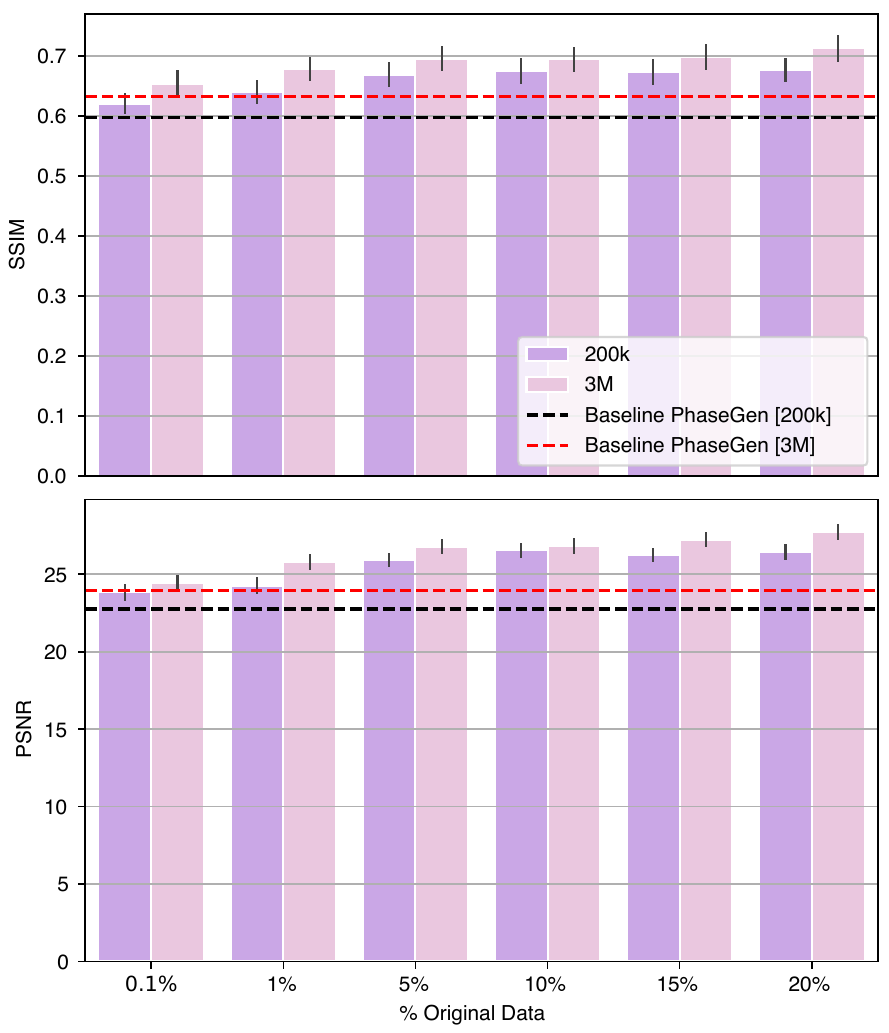}
    \caption{Results of the reconstruction task with different amounts of original-phase data. Shown are SSIM (top) and PSNR (bottom) for both the 200k model as well as the 3M model. The results are shown for an undersampling factor of four.}
    \label{fig:results_recon_ratio}
\end{figure}

Thus, in another experiment, the feasibility of using generated data by the proposed method in combination with original-phase data is explored. In the following experiment, we train the model with varying amounts of original-phase data. The model is trained with $0.1\%$, $1\%$, $5\%$, $10\%$, $15\%$ and $20\%$ of the available original-phase data, while the rest is generated with the proposed method. This experiment is conducted with an undersampling factor of four. The results are shown in Tab.\ref{tab:4} and Fig.\ref{fig:results_recon_ratio}.

\begin{table}[!tb]
    \caption{Results of the reconstruction task. Compared are different amounts of original-phase data. The results are shown for an undersampling factor of four.}
    \resizebox{\columnwidth}{!}{%
    \centering
    \def\arraystretch{1.2}
    \begin{tabular}{lllll}
    \toprule
    \multicolumn{1}{c}{\% of real worl data} & \multicolumn{1}{c}{SSIM ($\%) \uparrow$} & \multicolumn{1}{c}{PSNR (dB) $\uparrow$} & \multicolumn{1}{c}{MSE $\downarrow$} &  \multicolumn{1}{c}{NRMSE $\downarrow$}\\
    \midrule
    209k parameters & & & & \\
    \hspace{0.8cm} 0.1\% & 62.00 ± 9.92 & 23.84 ± 2.76 & 0.005 ± 0.004 & 0.310 ± 0.131 \\
    \hspace{0.8cm} 1\% & 63.98 ± 10.37 & 24.26 ± 27.15 & 0.005 ± 0.004 & 0.301 ± 0.143 \\
    \hspace{0.8cm} 5\% & 66.87 ± 10.90 & 25.95 ± 2.21 & 0.003 ± 0.002 & 0.240 ± 0.083 \\
    \hspace{0.8cm} 10\% & 67.42 ± 11.36 & 26.55 ± 2.47 & 0.003 ± 0.002 & 0.223 ± 0.071 \\
    \hspace{0.8cm} 15\% & 67.31 ± 11.07 & 26.22 ± 2.13 & 0.003 ± 0.001 & 0.231 ± 0.073 \\
    \hspace{0.8cm} 20\% & 67.62 ± 11.12 & 26.46 ± 2.61 & 0.003 ± 0.002 & 0.223 ± 0.058 \\
    \hspace{0.8cm} 100\% & 69.10 ± 11.44 & 27.19 ± 2.22 & 0.002 ± 0.001 & 0.207 ± 0.061 \\
    \midrule
    3.3M parameters & & & & \\    
    \hspace{0.8cm} 0.1\% & 65.34 ± 11.76 & 24.42 ± 3.02 & 0.005 ± 0.004 & 0.295 ± 0.135 \\
    \hspace{0.8cm} 1\% & 67.80 ± 10.78 & 25.78 ± 2.67 & 0.003 ± 0.003 & 0.249 ± 0.100 \\
    \hspace{0.8cm} 5\% & 69.52 ± 11.74 & 26.77 ± 2.55 & 0.003 ± 0.002 & 0.219 ± 0.007 \\
    \hspace{0.8cm} 10\% & 69.65 ± 11.66 & 26.90 ± 2.44 & 0.002 ± 0.002 & 0.218 ± 0.082 \\
    \hspace{0.8cm} 15\% & 69.78 ± 11.67 & 27.22 ± 2.43 & 0.002 ± 0.002 & 0.207 ± 0.070 \\
    \hspace{0.8cm} 20\% & 71.26 ± 12.06 & 27.72 ± 2.65 & 0.002 ± 0.001 & 0.198 ± 0.071 \\
    \hspace{0.8cm} 100\% & 71.28 ± 11.76 & 27.98 ± 2.58 & 0.002 ± 0.001 & 0.190 ± 0.066 \\
    \bottomrule
    \end{tabular}}
    \label{tab:4}
\end{table}

The 3M model trained with 15\% - 20\% of original-phase data achieves comparable results to the model trained with 100\% of original-phase data, while even the model trained with only 1\% of original-phase data increases by roughly 4.5\% points in PSNR and SSIM compared to the model trained with only synthetic data. 
To show the impact of the additional synthetic data, we conducted the same experiment with only the percentage of original-phase data without any additional generated data. With 10\% of original-phase data and no added synthetic data, the 209k model achieves a PSNR of $25.86 \pm 2.56$dB and a SSIM of $65.73 \pm 10.13\%$. When adding the generated data, the model achieves a PSNR of $26.55 \pm 2.47$dB and a SSIM of $67.42 \pm 11.36\%$. 
These results show that the proposed method can significantly reduce the amount of original phase data required for training, while achieving comparable results to a model trained only on original-phase data.

\section{Discussion}

The proposed complex-valued diffusion model shows promising results in generating synthetic phase data for MRI raw data. In the case of skullstripping, the model trained with solely generated raw data is able to generalize on original-phase data, while the model trained with no phase data or naive phase generation is not able to produce meaningful results. In the case of MRI reconstruction, the model trained with synthetic raw data generated by PhaseGen surpasses the naive phase generation. When training with a mix of real and synthetic data, the model achieves performance comparable to one trained entirely on original-phase data while using only 15–20\% real data. This shows the potential of the proposed method to significantly reduce the amount of original-phase data needed for training, while still achieving comparable results.

While this model is able to generate synthetic MRI raw data, there are still some limitations. The model is trained on single-coil data, while alot of data is commonly gahered with multiple coils. With an inference time of roughly 10 seconds per slice on a GPU, the generation of large datasets is still time consuming, especially when taking into account multiple coils. Future work will focus on the training of a model on multi-coil data, as well as a faster inference pipeline.
Another limitation is the lack of comparable datasets for further validation of the proposed method. While the skullstripping task shows the benefits of our model, we did not compare the results with a model trained on real MRI raw data, due to the lack of available datasets.

Interestingly, the naive method generates synthetic raw data that proves useful in certain use cases. While the model trained with naive phase generation is not able to reach the same results as the model trained with original-phase data, it is still able to produce meaningful results in the reconstruction task.

\section{Conclusion}

In this work, we introduce a complex-valued diffusion model capable of generating synthetic MRI raw data, guided by magnitude images. We demonstrate its effectiveness in generating synthetic phase data for skull stripping directly in k-Space. The model trained with synthetic data successfully generalized to original-phase data, whereas models trained without phase information or with naive phase generation failed to produce meaningful results. Additionally, we show that the generated data can be used to train a reconstruction model. The reconstruction model trained with synthetic data achieved performance comparable to one trained on original-phase data while requiring only 15–20\% real data.

In this study a neural network capable of generating complex-valued synthetic MRI raw data is presented, guided by magnitude image domain data. This publicly available model can be used by researchers to generate synthetic MRI raw data for pretraining their models, before fine-tuning them on original-phase data. This work will help to improve the research in the field of MRI raw data and its downstream tasks.

\section{Acknowledgement}

This work received funding from the the Bruno \& Helene Jöster Foundation, KITE (Plattform für KI-Translation Essen) from the REACT-EU initiative (https://kite.ikim.nrw/) and the Cancer Research
Center Cologne Essen (CCCE).
The authors acknowledge that this manuscript was edited with the assistance of LLMs. The authors declare no competing interests.

The following Figures have been created with BioRender:
\begin{itemize}
    \item Figure \ref{fig:overview}: Created in BioRender. Rempe, M. (2025) https://BioRender.com/z73n288
\end{itemize}

\section{References}

\printbibliography[heading=none]

\onecolumn
\include{appendix.tex}

\end{document}

%% file: appendix.tex
\section{Appendix}
\setcounter{table}{0}
\renewcommand{\thetable}{A\arabic{table}}
\setcounter{figure}{0}
\renewcommand{\thefigure}{A\arabic{figure}}

\subsection{Datasets}

\begin{table}[!htb]
    \caption{Overview of the used datasets.}
    \centering
    \def\arraystretch{1.6}
    \begin{tabular}{c|c|c|c|c}
    \hline
    {Dataset} & {Source} & {Type} & {Size} & {Magnetic Field}\\
    \hline
    PhaseGen Training \& Validation & University Hospital Essen & Raw MRI & 12071 slices & 1.5T / 3T \\
    Skullstrip Training \& Validation & University Hospital Essen & Image Domain & 21822 slices & 1.5T / 3T\\
    Skullstrip Testing & University Hospital Essen & Raw MRI & 14 volumes & 1.5T / 3T\\
    Reconstruction Training \& Validation & FastMRI & Raw Knee MRI & 40450 slices & 1.5T / 3T\\
    Reconstruction Testing & FastMRI & Raw Knee MRI & 1427 slices & 1.5T / 3T\\
    \hline
    \end{tabular}
    \label{tab:datasets}
\end{table}

\subsection{Hyperparameters}

\begin{table}[!htb]
    \caption{Hyperparameters for the skullstripping task.}
    \centering
    \def\arraystretch{1.6}
    \begin{tabular}{c|c}
    \hline
    {Hyperparameter} & {PhaseGen}\\
    \hline
    Learning Rate & 1e-4 \\
    Batch Size & 128 \\
    Dropout & 0.2 \\
    Epochs & 200 \\
    Noise Scheduler & Exponential \\
    Activation function & PReLU \\
    Optimizer & Adam \\
    \# Parameters & 33.5M \\
    \hline
    \end{tabular}
    \label{tab:hyperparameters_skullstrip}
\end{table}
\begin{table}[!htb]
    \caption{Hyperparameters for the reconstruction task.}
    \centering
    \def\arraystretch{1.6}
    \begin{tabular}{c|c|c|c}
    \hline
    {Hyperparameter} & {No Phase} & {Naive} & {PhaseGen}\\
    \hline
    Learning Rate & 3.8e-4 & 3.8e-4 & 4.4e-4\\
    Batch Size & 32 & 32 & 128\\
    Dropout & 0.1 & 0.1 & 0.4\\
    Epochs & 180 & 185 & 275\\
    Noise Scheduler & Exponential & Exponential & Exponential\\
    Activation function & ReLU & ReLU & ReLU\\
    Optimizer & AdamW & AdamW & AdamW\\
    \# Parameters & 209k / 3.3M & 209k / 3.3M & 209k / 3.3M\\
    \hline
    \end{tabular}
    \label{tab:hyperparameters_reconstruction}
\end{table}

%% file: lit.bib
@article{eggerDeepLearningFirst2021a,
  title = {Deep Learning---a First Meta-Survey of Selected Reviews across Scientific Disciplines, Their Commonalities, Challenges and Research Impact},
  author = {Egger, Jan and Pepe, Antonio and Gsaxner, Christina and Jin, Yuan and Li, Jianning and Kern, Roman},
  year = {2021},
  month = nov,
  journal = {PeerJ Computer Science},
  volume = {7},
  pages = {e773},
  issn = {2376-5992},
  doi = {10.7717/peerj-cs.773},
  copyright = {https://creativecommons.org/licenses/by/4.0/},
  langid = {english},
}

@misc{rempeTumorLikelihoodEstimation2024,
  title = {Tumor Likelihood Estimation on {{MRI}} Prostate Data by Utilizing K-{{Space}} Information},
  author = {Rempe, M. and H{\"o}rst, F. and Seibold, C. and Hadaschik, B. and Schlimbach, M. and Egger, J. and Kr{\"o}ninger, K. and Breuer, F. and Blaimer, M. and Kleesiek, J.},
  year = {2024},
  publisher = {arXiv},
  doi = {10.48550/ARXIV.2407.06165},
  copyright = {Creative Commons Attribution 4.0 International}
}

@incollection{liClassificationRegressionSegmentation2025,
  title = {Classification, {{Regression}} and {{Segmentation Directly}} from {{K-Space}} in {{Cardiac MRI}}},
  booktitle = {Machine {{Learning}} in {{Medical Imaging}}},
  author = {Li, Ruochen and Pan, Jiazhen and Zhu, Youxiang and Ni, Juncheng and Rueckert, Daniel},
  editor = {Xu, Xuanang and Cui, Zhiming and Rekik, Islem and Ouyang, Xi and Sun, Kaicong},
  year = {2025},
  volume = {15241},
  pages = {31--41},
  publisher = {Springer Nature Switzerland},
  address = {Cham},
  doi = {10.1007/978-3-031-73284-3_4},
  isbn = {978-3-031-73283-6 978-3-031-73284-3},
  langid = {english}
}

@article{dishnerSurveyPubliclyAvailable2024,
  title = {A {{Survey}} of {{Publicly Available}} {{{\textsc{MRI}}}} {{Datasets}} for {{Potential Use}} in {{Artificial Intelligence Research}}},
  shorttitle = {A {{Survey}} of {{Publicly Available}}},
  author = {Dishner, Katharine A. and McRae‐Posani, Bala and Bhowmik, Arka and Jochelson, Maxine S. and Holodny, Andrei and Pinker, Katja and Eskreis‐Winkler, Sarah and Stember, Joseph N.},
  date = {2024-02},
  journaltitle = {Journal of Magnetic Resonance Imaging},
  shortjournal = {Magnetic Resonance Imaging},
  volume = {59},
  number = {2},
  pages = {450--480},
  issn = {1053-1807, 1522-2586},
  doi = {10.1002/jmri.29101},
  url = {https://onlinelibrary.wiley.com/doi/10.1002/jmri.29101},
  langid = {english}
}

@article{singhEmergingTrendsFast2023,
  title = {Emerging {{Trends}} in {{Fast MRI Using Deep-Learning Reconstruction}} on {{Undersampled}} k-{{Space Data}}: {{A Systematic Review}}},
  shorttitle = {Emerging {{Trends}} in {{Fast MRI Using Deep-Learning Reconstruction}} on {{Undersampled}} k-{{Space Data}}},
  author = {Singh, Dilbag and Monga, Anmol and De Moura, Hector L. and Zhang, Xiaoxia and Zibetti, Marcelo V. W. and Regatte, Ravinder R.},
  year = {2023},
  month = aug,
  journal = {Bioengineering},
  volume = {10},
  number = {9},
  pages = {1012},
  issn = {2306-5354},
  doi = {10.3390/bioengineering10091012},
  copyright = {https://creativecommons.org/licenses/by/4.0/},
  langid = {english}
}

@article{zbontarFastMRIOpenDataset2018,
  title = {{{fastMRI}}: {{An Open Dataset}} and {{Benchmarks}} for {{Accelerated MRI}}},
  shorttitle = {{{fastMRI}}},
  author = {Zbontar, Jure and Knoll, Florian and Sriram, Anuroop and Murrell, Tullie and Huang, Zhengnan and Muckley, Matthew J. and Defazio, Aaron and Stern, Ruben and Johnson, Patricia and Bruno, Mary and Parente, Marc and Geras, Krzysztof J. and Katsnelson, Joe and Chandarana, Hersh and Zhang, Zizhao and Drozdzal, Michal and Romero, Adriana and Rabbat, Michael and Vincent, Pascal and Yakubova, Nafissa and Pinkerton, James and Wang, Duo and Owens, Erich and Zitnick, C. Lawrence and Recht, Michael P. and Sodickson, Daniel K. and Lui, Yvonne W.},
  date = {2018},
  publisher = {arXiv},
  doi = {10.48550/ARXIV.1811.08839},
  url = {https://arxiv.org/abs/1811.08839},
  version = {2}
}

@article{croitoruDiffusionModelsVision2023,
  title = {Diffusion {{Models}} in {{Vision}}: {{A Survey}}},
  shorttitle = {Diffusion {{Models}} in {{Vision}}},
  author = {Croitoru, Florinel-Alin and Hondru, Vlad and Ionescu, Radu Tudor and Shah, Mubarak},
  year = {2023},
  month = sep,
  journal = {IEEE Transactions on Pattern Analysis and Machine Intelligence},
  volume = {45},
  number = {9},
  pages = {10850--10869},
  issn = {0162-8828, 2160-9292, 1939-3539},
  doi = {10.1109/TPAMI.2023.3261988},
  copyright = {https://ieeexplore.ieee.org/Xplorehelp/downloads/license-information/IEEE.html}
}

@article{knollFastMRIPubliclyAvailable2020,
  title = {{{fastMRI}}: {{A Publicly Available Raw}} k-{{Space}} and {{DICOM Dataset}} of {{Knee Images}} for {{Accelerated MR Image Reconstruction Using Machine Learning}}},
  shorttitle = {{{fastMRI}}},
  author = {Knoll, Florian and Zbontar, Jure and Sriram, Anuroop and Muckley, Matthew J. and Bruno, Mary and Defazio, Aaron and Parente, Marc and Geras, Krzysztof J. and Katsnelson, Joe and Chandarana, Hersh and Zhang, Zizhao and Drozdzalv, Michal and Romero, Adriana and Rabbat, Michael and Vincent, Pascal and Pinkerton, James and Wang, Duo and Yakubova, Nafissa and Owens, Erich and Zitnick, C. Lawrence and Recht, Michael P. and Sodickson, Daniel K. and Lui, Yvonne W.},
  date = {2020-01-01},
  journaltitle = {Radiology: Artificial Intelligence},
  shortjournal = {Radiology: Artificial Intelligence},
  volume = {2},
  number = {1},
  pages = {e190007},
  issn = {2638-6100},
  doi = {10.1148/ryai.2020190007},
  url = {http://pubs.rsna.org/doi/10.1148/ryai.2020190007},
  langid = {english}
}

@online{tibrewalaFastMRIProstatePublicly2023,
  title = {{{FastMRI Prostate}}: {{A Publicly Available}}, {{Biparametric MRI Dataset}} to {{Advance Machine Learning}} for {{Prostate Cancer Imaging}}},
  shorttitle = {{{FastMRI Prostate}}},
  author = {Tibrewala, Radhika and Dutt, Tarun and Tong, Angela and Ginocchio, Luke and Keerthivasan, Mahesh B and Baete, Steven H and Chopra, Sumit and Lui, Yvonne W and Sodickson, Daniel K and Chandarana, Hersh and Johnson, Patricia M},
  date = {2023},
  doi = {10.48550/ARXIV.2304.09254},
  url = {https://arxiv.org/abs/2304.09254},
  pubstate = {prepublished},
  version = {1}
}

@article{heckelDeepLearningAccelerated2024,
  title = {Deep Learning for Accelerated and Robust {{MRI}} Reconstruction},
  author = {Heckel, Reinhard and Jacob, Mathews and Chaudhari, Akshay and Perlman, Or and Shimron, Efrat},
  date = {2024-07-23},
  journaltitle = {Magnetic Resonance Materials in Physics, Biology and Medicine},
  shortjournal = {Magn Reson Mater Phy},
  volume = {37},
  number = {3},
  pages = {335--368},
  issn = {1352-8661},
  doi = {10.1007/s10334-024-01173-8},
  url = {https://link.springer.com/10.1007/s10334-024-01173-8},
  langid = {english},
}

@article{rempeKstripNovelSegmentation2024a,
  title = {K-Strip: {{A}} Novel Segmentation Algorithm in k-Space for the Application of Skull Stripping},
  shorttitle = {K-Strip},
  author = {Rempe, Moritz and Mentzel, Florian and Pomykala, Kelsey L. and Haubold, Johannes and Nensa, Felix and Kroeninger, Kevin and Egger, Jan and Kleesiek, Jens},
  date = {2024-01},
  journaltitle = {Computer Methods and Programs in Biomedicine},
  shortjournal = {Computer Methods and Programs in Biomedicine},
  volume = {243},
  pages = {107912},
  issn = {01692607},
  doi = {10.1016/j.cmpb.2023.107912},
  url = {https://linkinghub.elsevier.com/retrieve/pii/S0169260723005783},
  urldate = {2025-01-20},
  langid = {english}
}

@article{leeComplexValuedNeuralNetworks2022,
  title = {Complex-{{Valued Neural Networks}}: {{A Comprehensive Survey}}},
  shorttitle = {Complex-{{Valued Neural Networks}}},
  author = {Lee, ChiYan and Hasegawa, Hideyuki and Gao, Shangce},
  date = {2022-08},
  journaltitle = {IEEE/CAA Journal of Automatica Sinica},
  shortjournal = {IEEE/CAA J. Autom. Sinica},
  volume = {9},
  number = {8},
  pages = {1406--1426},
  issn = {2329-9266, 2329-9274},
  doi = {10.1109/JAS.2022.105743},
  url = {https://ieeexplore.ieee.org/document/9849162/},
  urldate = {2025-01-20}
}

@online{smithComplexValuedNeuralNetworks2023,
  title = {Complex-{{Valued Neural Networks}} for {{Data-Driven Signal Processing}} and {{Signal Understanding}}},
  author = {Smith, Josiah W.},
  date = {2023},
  doi = {10.48550/ARXIV.2309.07948},
  url = {https://arxiv.org/abs/2309.07948},
  pubstate = {prepublished},
  version = {1}
}

@inproceedings{hayakawaApplyingComplexValuedNeural2018,
  title = {Applying {{Complex-Valued Neural Networks}} to {{Acoustic Modeling}} for {{Speech Recognition}}},
  booktitle = {2018 {{Asia-Pacific Signal}} and {{Information Processing Association Annual Summit}} and {{Conference}} ({{APSIPA ASC}})},
  author = {Hayakawa, Daichi and Masuko, Takashi and Fujimura, Hiroshi},
  date = {2018-11},
  pages = {1725--1731},
  publisher = {IEEE},
  location = {Honolulu, HI, USA},
  doi = {10.23919/APSIPA.2018.8659610},
  url = {https://ieeexplore.ieee.org/document/8659610/},
  eventtitle = {2018 {{Asia-Pacific Signal}} and {{Information Processing Association Annual Summit}} and {{Conference}} ({{APSIPA ASC}})},
  isbn = {978-988-14768-5-2}
}

@article{shortenSurveyImageData2019,
  title = {A Survey on {{Image Data Augmentation}} for {{Deep Learning}}},
  author = {Shorten, Connor and Khoshgoftaar, Taghi M.},
  date = {2019-12},
  journaltitle = {Journal of Big Data},
  shortjournal = {J Big Data},
  volume = {6},
  number = {1},
  pages = {60},
  issn = {2196-1115},
  doi = {10.1186/s40537-019-0197-0},
  url = {https://journalofbigdata.springeropen.com/articles/10.1186/s40537-019-0197-0},
  langid = {english}
}

@online{kingmaAutoEncodingVariationalBayes2013,
  title = {Auto-{{Encoding Variational Bayes}}},
  author = {Kingma, Diederik P and Welling, Max},
  date = {2013},
  doi = {10.48550/ARXIV.1312.6114},
  url = {https://arxiv.org/abs/1312.6114},
  pubstate = {prepublished},
  version = {11}
}

@article{goodfellowGenerativeAdversarialNetworks2020,
  title = {Generative Adversarial Networks},
  author = {Goodfellow, Ian and Pouget-Abadie, Jean and Mirza, Mehdi and Xu, Bing and Warde-Farley, David and Ozair, Sherjil and Courville, Aaron and Bengio, Yoshua},
  date = {2020-10-22},
  journaltitle = {Communications of the ACM},
  shortjournal = {Commun. ACM},
  volume = {63},
  number = {11},
  pages = {139--144},
  issn = {0001-0782, 1557-7317},
  doi = {10.1145/3422622},
  url = {https://dl.acm.org/doi/10.1145/3422622},
  langid = {english}
}

@online{ferreiraImprovedMultiTaskBrain2024,
  title = {Improved {{Multi-Task Brain Tumour Segmentation}} with {{Synthetic Data Augmentation}}},
  author = {Ferreira, André and Jesus, Tiago and Puladi, Behrus and Kleesiek, Jens and Alves, Victor and Egger, Jan},
  date = {2024-12-02},
  eprint = {2411.04632},
  eprinttype = {arXiv},
  eprintclass = {cs},
  doi = {10.48550/arXiv.2411.04632},
  url = {http://arxiv.org/abs/2411.04632},
  urldate = {2025-01-20},
  pubstate = {prepublished}
}

@online{trabuccoEffectiveDataAugmentation2023,
  title = {Effective {{Data Augmentation With Diffusion Models}}},
  author = {Trabucco, Brandon and Doherty, Kyle and Gurinas, Max and Salakhutdinov, Ruslan},
  date = {2023-05-25},
  eprint = {2302.07944},
  eprinttype = {arXiv},
  eprintclass = {cs},
  doi = {10.48550/arXiv.2302.07944},
  url = {http://arxiv.org/abs/2302.07944},
  pubstate = {prepublished}
}

@incollection{korkmazSelfsupervisedMRIReconstruction2023,
  title = {Self-Supervised {{MRI Reconstruction}} with {{Unrolled Diffusion Models}}},
  booktitle = {Medical {{Image Computing}} and {{Computer Assisted Intervention}} – {{MICCAI}} 2023},
  author = {Korkmaz, Yilmaz and Cukur, Tolga and Patel, Vishal M.},
  editor = {Greenspan, Hayit and Madabhushi, Anant and Mousavi, Parvin and Salcudean, Septimiu and Duncan, James and Syeda-Mahmood, Tanveer and Taylor, Russell},
  date = {2023},
  volume = {14229},
  pages = {491--501},
  publisher = {Springer Nature Switzerland},
  location = {Cham},
  doi = {10.1007/978-3-031-43999-5_47},
  url = {https://link.springer.com/10.1007/978-3-031-43999-5_47},
  isbn = {978-3-031-43998-8 978-3-031-43999-5},
  langid = {english}
}

@online{nachmaniNonGaussianDenoising2021,
  title = {Non {{Gaussian Denoising Diffusion Models}}},
  author = {Nachmani, Eliya and Roman, Robin San and Wolf, Lior},
  date = {2021},
  doi = {10.48550/ARXIV.2106.07582},
  url = {https://arxiv.org/abs/2106.07582},
  pubstate = {prepublished},
  version = {1}
}

@online{hoDenoisingDiffusionProbabilistic2020a,
  title = {Denoising {{Diffusion Probabilistic Models}}},
  author = {Ho, Jonathan and Jain, Ajay and Abbeel, Pieter},
  date = {2020},
  doi = {10.48550/ARXIV.2006.11239},
  url = {https://arxiv.org/abs/2006.11239},
  pubstate = {prepublished},
  version = {2}
}

@article{bradburyMesentericVenousThrombosis2002,
  title = {Mesenteric {{Venous Thrombosis}}: {{Diagnosis}} and {{Noninvasive Imaging}}},
  shorttitle = {Mesenteric {{Venous Thrombosis}}},
  author = {Bradbury, Michelle S. and Kavanagh, Peter V. and Bechtold, Robert E. and Chen, Michael Y. and Ott, David J. and Regan, John D. and Weber, Therese M.},
  date = {2002-05},
  journaltitle = {RadioGraphics},
  shortjournal = {RadioGraphics},
  volume = {22},
  number = {3},
  pages = {527--541},
  issn = {0271-5333, 1527-1323},
  doi = {10.1148/radiographics.22.3.g02ma10527},
  url = {http://pubs.rsna.org/doi/10.1148/radiographics.22.3.g02ma10527},
  langid = {english}
}

@article{wuIdentificationCalcificationMRI2009,
  title = {Identification of Calcification with {{MRI}} Using Susceptibility‐weighted Imaging: {{A}} Case Study},
  shorttitle = {Identification of Calcification with {{MRI}} Using Susceptibility‐weighted Imaging},
  author = {Wu, Zhen and Mittal, Sandeep and Kish, Karl and Yu, Yingjian and Hu, J. and Haacke, E. Mark},
  date = {2009-01},
  journaltitle = {Journal of Magnetic Resonance Imaging},
  shortjournal = {Magnetic Resonance Imaging},
  volume = {29},
  number = {1},
  pages = {177--182},
  issn = {1053-1807, 1522-2586},
  doi = {10.1002/jmri.21617},
  url = {https://onlinelibrary.wiley.com/doi/10.1002/jmri.21617},
  langid = {english}
}

@article{warfieldSimultaneousTruthPerformance2004a,
  title = {Simultaneous {{Truth}} and {{Performance Level Estimation}} ({{STAPLE}}): {{An Algorithm}} for the {{Validation}} of {{Image Segmentation}}},
  shorttitle = {Simultaneous {{Truth}} and {{Performance Level Estimation}} ({{STAPLE}})},
  author = {Warfield, S.K. and Zou, K.H. and Wells, W.M.},
  date = {2004-07},
  journaltitle = {IEEE Transactions on Medical Imaging},
  shortjournal = {IEEE Trans. Med. Imaging},
  volume = {23},
  number = {7},
  pages = {903--921},
  issn = {0278-0062},
  doi = {10.1109/TMI.2004.828354},
  url = {http://ieeexplore.ieee.org/document/1309714/},
  langid = {english}
}

@article{isenseeAutomatedBrainExtraction2019b,
  title = {Automated Brain Extraction of Multisequence {{MRI}} Using Artificial Neural Networks},
  author = {Isensee, Fabian and Schell, Marianne and Pflueger, Irada and Brugnara, Gianluca and Bonekamp, David and Neuberger, Ulf and Wick, Antje and Schlemmer, Heinz‐Peter and Heiland, Sabine and Wick, Wolfgang and Bendszus, Martin and Maier‐Hein, Klaus H. and Kickingereder, Philipp},
  date = {2019-12},
  journaltitle = {Human Brain Mapping},
  shortjournal = {Human Brain Mapping},
  volume = {40},
  number = {17},
  pages = {4952--4964},
  issn = {1065-9471, 1097-0193},
  doi = {10.1002/hbm.24750},
  url = {https://onlinelibrary.wiley.com/doi/10.1002/hbm.24750},
  langid = {english}
}

@article{hoopesSynthStripSkullstrippingAny2022a,
  title = {{{SynthStrip}}: Skull-Stripping for Any Brain Image},
  shorttitle = {{{SynthStrip}}},
  author = {Hoopes, Andrew and Mora, Jocelyn S. and Dalca, Adrian V. and Fischl, Bruce and Hoffmann, Malte},
  date = {2022-10},
  journaltitle = {NeuroImage},
  shortjournal = {NeuroImage},
  volume = {260},
  pages = {119474},
  issn = {10538119},
  doi = {10.1016/j.neuroimage.2022.119474},
  url = {https://linkinghub.elsevier.com/retrieve/pii/S1053811922005900},
  langid = {english}
}

@online{rempeDeIdentificationMedicalImaging2024,
  title = {De-{{Identification}} of {{Medical Imaging Data}}: {{A Comprehensive Tool}} for {{Ensuring Patient Privacy}}},
  shorttitle = {De-{{Identification}} of {{Medical Imaging Data}}},
  author = {Rempe, Moritz and Heine, Lukas and Seibold, Constantin and Hörst, Fabian and Kleesiek, Jens},
  date = {2024},
  doi = {10.48550/ARXIV.2410.12402},
  url = {https://arxiv.org/abs/2410.12402},
  pubstate = {prepublished},
  version = {1}
}

@article{dewey2022laplacian,
  title={Laplacian-Based Phase Unwrapping in Python (v1. 0)},
  author={Dewey, Blake E},
  journal={Zenodo. https://doi. org/10.5281/zenodo},
  volume={7198991},
  year={2022}
}

@article{hirose2006complex,
  title={Complex-valued neural networks},
  author={Hirose, Akira and others},
  volume={32},
  year={2006},
  publisher={Wiley Online Library}
}

@article{hyun2018deep,
  title={Deep learning for undersampled MRI reconstruction},
  author={Hyun, Chang Min and Kim, Hwa Pyung and Lee, Sung Min and Lee, Sungchul and Seo, Jin Keun},
  journal={Physics in Medicine \& Biology},
  volume={63},
  number={13},
  pages={135007},
  year={2018},
  publisher={IOP Publishing}
}

@article{egger2022medical,
  title={Medical deep learning—A systematic meta-review},
  author={Egger, Jan and Gsaxner, Christina and Pepe, Antonio and Pomykala, Kelsey L and Jonske, Frederic and Kurz, Manuel and Li, Jianning and Kleesiek, Jens},
  journal={Computer methods and programs in biomedicine},
  volume={221},
  pages={106874},
  year={2022},
  publisher={Elsevier}
}

@article{wang2025deep,
  title={Deep learning on medical image analysis},
  author={Wang, Jiaji and Wang, Shuihua and Zhang, Yudong},
  journal={CAAI Transactions on Intelligence Technology},
  volume={10},
  number={1},
  pages={1--35},
  year={2025},
  publisher={Wiley Online Library}
}

@article{albuquerque2025deep,
  title={Deep learning-based object detection algorithms in medical imaging: Systematic review},
  author={Albuquerque, Carina and Henriques, Roberto and Castelli, Mauro},
  journal={Heliyon},
  volume={11},
  number={1},
  year={2025},
  publisher={Elsevier}
}

@article{tibrewala2024fastmri,
  title={FastMRI Prostate: A public, biparametric MRI dataset to advance machine learning for prostate cancer imaging},
  author={Tibrewala, Radhika and Dutt, Tarun and Tong, Angela and Ginocchio, Luke and Lattanzi, Riccardo and Keerthivasan, Mahesh B and Baete, Steven H and Chopra, Sumit and Lui, Yvonne W and Sodickson, Daniel K and others},
  journal={Scientific Data},
  volume={11},
  number={1},
  pages={404},
  year={2024},
  publisher={Nature Publishing Group UK London}
}

@article{solomon2025fastmri,
  title={fastMRI Breast: A publicly available radial k-space dataset of breast dynamic contrast-enhanced MRI},
  author={Solomon, Eddy and Johnson, Patricia M and Tan, Zhengguo and Tibrewala, Radhika and Lui, Yvonne W and Knoll, Florian and Moy, Linda and Kim, Sungheon Gene and Heacock, Laura},
  journal={Radiology: Artificial Intelligence},
  volume={7},
  number={1},
  pages={e240345},
  year={2025},
  publisher={Radiological Society of North America}
}

@article{weng2020black,
  title={Black dipole or white dipole: Using susceptibility phase imaging to differentiate cerebral microbleeds from intracranial calcifications},
  author={Weng, C-L and Jeng, Y and Li, Y-T and Chen, C-J and Chen, DY-T},
  journal={American Journal of Neuroradiology},
  volume={41},
  number={8},
  pages={1405--1413},
  year={2020},
  publisher={American Journal of Neuroradiology}
}
